\documentclass[pra,twocolumn,showpacs,eqsecnum,superscriptaddress]{revtex4}
\usepackage{makeidx}
\usepackage{eurosym}
\usepackage{amssymb}
\usepackage{amsfonts}
\usepackage{amsmath}
\usepackage{setspace}
\usepackage[english]{babel}
\usepackage{graphicx}
\usepackage[applemac]{inputenc}
\usepackage{color}
\usepackage{ifthen}
\usepackage{float}
\usepackage{verbatim}
\usepackage{subfigure}
\begin{document}

\title{The ergodic and non-ergodic phases in one dimensional clean Jaynes-Cummings-Hubbard system}

\author{Jin-Lou Ma}
\affiliation{Lanzhou Center for Theoretical Physics, Key Laboratory of Theoretical Physics of Gansu Province, Lanzhou University, Lanzhou, Gansu $730000$, China}
\author{Qing Li}
\affiliation{Lanzhou Center for Theoretical Physics, Key Laboratory of Theoretical Physics of Gansu Province, Lanzhou University, Lanzhou, Gansu $730000$, China}
\author{Lei Tan}
\email{tanlei@lzu.edu.cn}
\affiliation{Lanzhou Center for Theoretical Physics, Key Laboratory of Theoretical Physics of Gansu Province, Lanzhou University, Lanzhou, Gansu $730000$, China}
\affiliation{Key Laboratory for Magnetism and Magnetic Materials of the Ministry of Education, Lanzhou University, Lanzhou $730000$, China}

\date{\today}
\begin{abstract}
We study the ergodic and non-ergodic behaviors of a clean Jaynes-Cummings-Hubbard chain for different parameters based on the average level spacings and the generalized fractal dimensions of eigenstates by using exact diagonalization. It can be found that a transition from ergodicity  to non-ergodicity phases happens when the atom-photon detuning is large, and the non-ergodic phases maybe exist in the thermodynamic limit. We also find that the non-ergodic phase violates the eigenstate thermalization hypothesis.
Finally, we study the many-body multifractality of the ground state and find that the derivative of the generalized fractal dimensions can determine the critical point of the Superfluid-Mott-insulation phase transition in a small range of parameters under different boundary conditions and there is no ergodicity for the ground state.
\end{abstract}


\maketitle

\section{Introduction}
Generally speaking, ergodicity is the basic hypothesis of classical statistical physics, which means that the system will visit each region of phase space in the process of time evolution and the time spent in each region is proportional to the volume of the system under the long-term limit \cite{Ozorio,Ott,LDAlessio}. Learning from the idea of classical statistical mechanics, how to connect quantum mechanics with statistical physics has become a topic of interest. Based on Berry-Tabor conjecture\cite{MVBerry}, quantum ergodic theory comes into being: for any quantum system, it is assumed that the eigenstate of ergodic Hamiltonian is essentially an uncorrelated random variable\cite{JMDeutsch}. Therefore, using the random matrix theory, the eigenvalues and eigenstates of many-body quantum systems can predict the existence of ergodicity\cite{Berry,Delande,Khaymovich,Tomasi,DAlessio,Haake,Chirikov}.

It was realized very early that not all systems are ergodicity, and some quantum integrable systems are non-ergodic. When a generic perturbation is added, a few quantum integrable systems are unstable and become ergodic\cite{Deutsch}. Ergodicity is an obvious property of quantum nonintegrable systems, but ergodicity breaking has been found in many cases due to different factors. For example, for a classical periodic driven one degree of freedom\cite{Chirikov} or many-body systems\cite{Rozenbaum,Rylands,Fava,Notarnicola}, under the quantum regime, the interference effect leads to an inhibition of energy absorption and forms dynamic localization, which shows a non-ergodic behavior. When a disorder is introduced, a complex many-body system cannot be thermalized and produces a spatially localized motion integral with a non-ergodic behavior\cite{Abanin,Nandkishore,Imbrie}. Since 2017, researchers have found that in lattice gauge theories, the emergence of local constraints leads to non-ergodicity\cite{Smith,Smith1,Brenes,Smith2,Smith3,Russomanno,Karpov}. Later, it was found that the large tilt potential can also produce a similar effect to the disorder, resulting in the
non-ergodic many-body systems\cite{Nieuwenburg,Schulz,Prosen,Prosen1}, which has been realized in the latest experiments\cite{Morong}. In addition, the existence of non-ergodic behaviors is also found for a one-dimensional uniform Josephson junction chain at higher energies or weak Josephson coupling\cite{Pino} and a clean Bose Hubbard chain under weak tunneling strength\cite{Russomanno1}. Interesting questions thus arise and need to be clarified, e.g., whether there are other factors that can lead to the ergodicity breaking?

Inspired by the above works, we are interested in exploring whether a clean Jaynes-Cummings-Hubbard (JCH) has a strict ergodicity under the thermodynamic limit, and whether the transition from ergodic to non-ergodic phase can be controlled by changing the detunings. JCH model describes a Boson and spin-like (two-level atom) hybrid system in quantum optics. In the study of ergodicity and non-ergodicity of many-body systems mentioned above, most researchers mainly focus on spin, Fermion or Boson systems with direct two-body interaction. While the ergodicity of the JCH model is formed by the competition between the effective on-site repulsive potential, caused by the atom-photon coupling, and photon tunneling strength. There are relatively few investigations on the ergodicity and thermalization based on the JCH model. Only our group has published relevant papers for the resonant cases\cite{qli,qli1}. But the influence of the atom-photon detuning is absent. As is well known, detuning is an important experimental parameter, which definitely affects the properties of the system. A large number of novel physics emerging from the detuning has been widely explored, which includes the photon blockade\cite{Werner}, non-Markovian behavior\cite{CYChen}, polariton Mott phases\cite{Aichhorn}, geometric quantum computation\cite{JGLi}, population transfer\cite{Stefano},  et al. Inspired by these works, in this paper, we focus on the non-ergodic and ergodic phase of one dimensional JCH model for different atom-photon detunings under almost all range of tunnelling strength based on the average level spacing ratios and the generalized fractal dimensions (GFDs). We find that the atom-photon detuning can control the ergodic and non-ergodic phase transition of the JCH model. With the increase of the absolute value of atom-photon detuning, the ergodic region of the system gradually decreases and then disappears. In the case of large detuning, through finite-size analysis, it is predicted that ergodic breaking also exists under the thermodynamic limit. Then, through the eigenstate thermalization hypothesis (ETH), we further know that the non-ergodic phase does not satisfy the thermalization behavior. The positive and negative atom-photon detuning have different effects on the ground state ergodicity. Under the exact diagonalization, it is found that the critical phase transition point of Superfluid (SF) to Mott-insulation (MI) can be determined in a very small range of parameters by the derivative of GFDs.

The paper is organized as follows: In Sec. \uppercase\expandafter{\romannumeral2} , the theoretical model is introduced, and we describe the required physical quantities and analyze the energy spectrum phase in Sec. \uppercase\expandafter{\romannumeral3}. Sec. \uppercase\expandafter{\romannumeral4} is devoted to providing the relationship between the ground state phase transition and multifractal. The conclusions are summarized in Sec. \uppercase\expandafter{\romannumeral5}.

\section{theoretical model}

We study a one dimensional JCH model including an atom-photon interaction term and a photon tunnelling term between nearest neighboring cavities\cite{Greentree,Angelakis,Makin}. After using the rotating transformation operator $U=exp[-i\sum^L_{j=1}\omega_c( a_j^{\dagger}a_j+\sigma^{+}_{j}\sigma^{-}_{j})t]$, the Hamiltonian can be written as
\begin{align}
\begin{split}
H=& H_{int}+H_{tun}\\
=& \sum^L_{i}[\Delta\sigma^{+}_{i}\sigma^{-}_{i}+g_a(a_i\sigma^{+}_{i}+a_i^{\dagger}\sigma^{-}_{i})]{}\\ &{}-t\sum^L_i (a_i^{\dagger}a_{i+1}+a_ia^{\dagger}_{i+1}),\label{H1}
\end{split}
\end{align}
where $\Delta=\omega_a-\omega_c$ is the atom-photon detuning, $\omega_a$ is the transition energy of the two-level atom in every cavity and $\omega_c$ is the frequency of each cavity field. $\sigma_i^{+}$ and $\sigma_i^{-}$ are the atomic raising and lowering operators, respectively. The second term of the local interaction Hamiltonian describes the on-site coupling between the photons and the atom on each site, and $a_i^{\dagger} (a_i)$ is the photonic creation (annihilation) operator in the $i$th site. We assume that all the atoms couple to cavities with the same coupling $g_a$. The tunnelling Hamiltonian term is the sum of tunnelling of photons and $t$ is the hopping energy of photons between the nearest neighboring cavities $i$ and $i+1$ for all cavities. This system includes $N$ excitations [$N=\sum_{i}(a_i^{\dagger}a_i+\sigma^{+}_{i}\sigma^{-}_{i})=\sum_{i}(n_i^{c}+n_i^{a})$ is the total number of atomic and photonic excitations]. The filling factor is $\nu \equiv N/L=1$, unless expressly specified otherwise. The JCH model has a reflection symmetry under the reflection (parity) operation ($P$) about the center cavity. Hilbert space can be decomposed into symmetric $p=1$ and antisymmetric $p=-1$ subspaces. In the periodic boundary condition (PBC), the JCH model also has translational symmetry and Hilbert space is further divided into subspaces with different quasimomentum $Q$. The general basis of full Hilbert space of $H$, with a space dimension,
\begin{align}
\mathcal{N}=\sum_{s=1}^{min[N,L]}(L\frac{(N+L-s-1)!}{(N-s)!(L-s)!s!}),
\end{align}
 is shown by the direct product of the cavity
field states and atom states $|\mathbf{n}\rangle\equiv \prod_{i}|n,e(g)\rangle_i$. We mainly use numerical calculation in the irreducible Hilbert subspace $\mathcal{D}\approx\mathcal{N}/2L$ with $Q=0$ and $p=-1$ for the PBC.

When one only considers the interaction term $H_{int}$ of the Hamiltonian, each on-site decouples and the effective local Jaynes-Cummings Hamiltonian can be easily diagonalized. One can obtain the eigenstates known as dressed states,
$|\pm,n_i\rangle= [(\Delta\pm\chi_{n_i})/2|n_i,g\rangle + g_a\sqrt{n_i}|n_i-1,e\rangle]/\sqrt{\chi_{n_i}\mp\chi_{n_i}/2}$ in the $i$th cavity, and the corresponding eigenenergies are $E_i^{\pm}=(\Delta\pm \chi_{n_i})/2$, with $\chi_{n_i}=\sqrt{4 g_a^2n_i+\Delta^2}$.
The eigenvectors of the local interaction Hamiltonian $H_{int}$ are $\prod_{i=1}^{L}|\pm,n_i\rangle$. The eigenvectors of the tunnelling Hamiltonian $H_{tun}$  are the delocalized plane-wave (standing-wave) modes with different wave vectors from the Fock states for PBC (hard-wall boundary condition).  The interaction term and the tunnelling term are integrable and analytically solvable in the real and momentum spaces, respectively. When the scaled tunnelling strength $t/g_a$ is around $1$, the competition between tunnelling and interaction makes the JCH model be nonintegrable and show spectral chaos and ergodicity\cite{qli1}. When $t/g_a\rightarrow 0$ ($\infty$), the JCH model exhibits integrability and non-ergodicity.

\begin{figure*}[htb]
\includegraphics[scale=0.9,trim=30 0 0 0]
{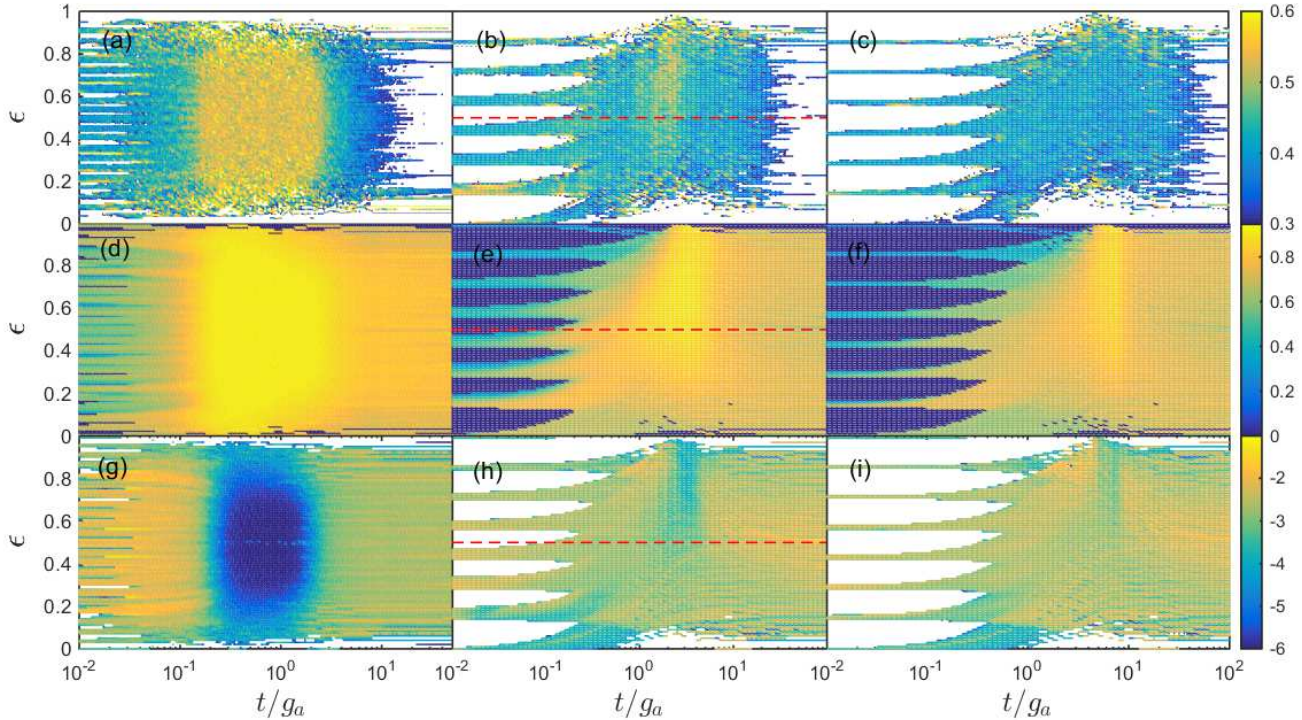}
\caption{(Color online)  The average level spacings $\langle r\rangle$ (top), the GFDs $\langle \widetilde{D}_{1}\rangle$ (middle) and the variance $log[var(\widetilde{D}_{1})]$ (bottom) as functions of the scaled tunnelling strength $t/g_a$ and the scaled energy $\epsilon$ with $L=8$, $\mathcal{D}=9581$  for PBC.  The first, second and third column correspond to $|\Delta/g_a|$ equaling to $0$, $5$ and $10$, respectively. Red dashed lines mark $\epsilon =0.5$. The white area indicates that it goes beyond the display range of $\langle r\rangle$ and $var(\widetilde{D}_{1})$.}\label{fig1}
\end{figure*}
\section{energy spectrum phase transition and thermalization behavior}
For characterizing the ergodicity of the JCH model from the eigenstate structure, the conventional method is based on the multifractal complexity by using finite-size GFDs\cite{Rodriguez,Lindinger},
\begin{align}
\widetilde{D}_q=\frac{1}{1-q}log_{\mathcal{N}}R_q, \quad with \quad  R_q=\sum_{\alpha}|\psi_{\alpha}|^{2q},q\in \mathbb{R}^+,\label{H2}
\end{align}
where $\psi_{\alpha}$ is the amplitude of eigenvectors in a given orthonormal basis of size $\mathcal{N}$. The scaling of $R_q$ is generically described by the from $R_q\sim \mathcal{N}^{-(q-1)D_q}$ with $D_q\equiv lim_{\mathcal{N}\rightarrow\infty}\widetilde{D}_q$ $(D_q\in[0, 1])$. When $\widetilde{D}_{q\geq1}=0$, the state is localized in the considered basis. When $\widetilde{D}_{q}=1$ for all $q$ moments, the state is ergodic. Otherwise, the state is extended non-ergodic (multifractal). In this paper, we only study the case $q=1,2$ and $\infty$. By using the $L'H\hat{o}pital$'s rule, we can get the dimension $D_1=-\sum_{\alpha}|\psi_{\alpha}|^2log_{\mathcal{N}}|\psi_{\alpha}|^2$, which is the information dimension and governs the scaling of the Shannon information entropy. $D_2$ is known as a measure of the state's "volume" and affects the inverse participation ratio of the eigenstate. As for $\widetilde{D}_{\infty}$, it is equal to $lim_{q\rightarrow\infty}log_{\mathcal{N}}[(\sum_{\alpha}|\psi_{\alpha}|^{2q})^{1/(q-1)}]\approx-log_{\mathcal{N}}(|\psi_\alpha|_{max[\alpha]}^2)$ , only the largest term $|\psi_\alpha|^2$ will contribute to the summation and $q/(q-1)\rightarrow 1$ in the limit $q\rightarrow\infty$, and determines the maximum value of the intensities in a certain basis\cite{Lindinger1}.

In this section, we analyze the phase transition of the energy spectrum for the ergodic and non-ergodic regimes for different $\Delta/g_a$. Eigenenergy is scaled as $\epsilon\equiv(E-E_{min})/(E_{max}-E_{min})\in[0,1]$, then the eigenstates can be chosen around energy targets\cite{Pietracaprina,Balay,Hernandez}. In the random matrix theory, the discrimination between ergodic and non-ergodic regimes can depend on spectral statistics\cite{Russomanno1,Mehta}. We can capture the statistical features of spectrum by the level spacing ratios\cite{Oganesyan,Atas}, $r_n=min(\delta_{n+1}/\delta_n,\delta_n/\delta_{n+1})$, where $\delta_n=E_{n+1}-E_n$ is the $n$th level spacing and the eigenenergies are arranged in an ascending order. When the system is ergodic, which is expressed as a Gaussian-orthogonal-ensemble (GOE) random matrix and the average level spacings show the Wigner-Dyson distribution ($\langle r\rangle_{WD}\approx0.5295$)\cite{Atas}. Otherwise, the system is non-ergodic, which is just like a classical integrable system and shows as Poisson distribution ($\langle r\rangle_{P}\approx0.386$)\cite{Berry1}. Next, combining $\langle r\rangle$ and $\langle \widetilde{D}_{1}\rangle$, we can judge which phase of the system is shown under different parameters. Due to the facts that the fluctuations of GFDs are particularly sensitive to ergodic behavior than the GFDs, we also consider its standard variance $var(\widetilde{D}_{1})$\cite{Pausch}. When the value of $var(\widetilde{D}_{1})$ changes with the extreme drop, the state undergoes a transition from nonergodic to ergodic. In order to elucidate the asymptotic behavior of GFDs in ergodic region, the numerical results of $\langle \widetilde{D}_{1}\rangle$ and $var(\widetilde{D}_{1})$ are compared against the GOE values for the random matrix theory (RMT), which provides wonderful analytical approximation\cite{Haake,Pausch},
\begin{align}
\langle \widetilde{D}_{1}\rangle_{GOE}=&\frac{H_{\frac{\mathcal{D}}{2}}-2+ln4}{ln\mathcal{D}},  \label{H3}\\ var(\widetilde{D}_{1})_{GOE}=&\frac{(3\pi^2-24)(\mathcal{D}+2)-8}{2(\mathcal{D}+2)^2ln^2\mathcal{D}}
-\frac{\psi^{(1)}(2+\frac{\mathcal{D}}{2})}{ln^2\mathcal{D}},\label{H4}
\end{align}
where $H_n=\sum^n_{k=1}\frac{1}{k}$ and $\psi^{(1)}$ expresses the first derivative of the digamma function\cite{Olver}.

\begin{figure}[H]
\centering
\includegraphics[height=9cm,width=9.8cm]{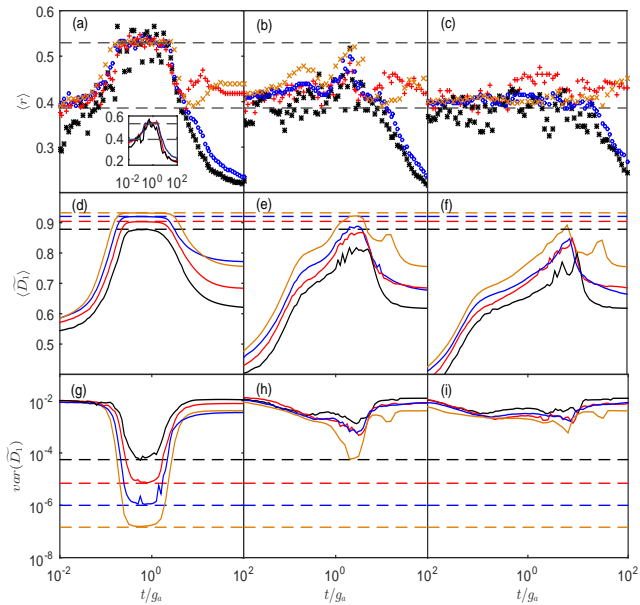}
\caption{(Color online) The average level spacings $\langle r\rangle$ (top), the GFDs $\langle \widetilde{D}_{1}\rangle$ (middle) and the standard variance $var(\widetilde{D}_{1})$ (bottom) versus $t/g_a$ at the middle one-third spectrum for different atom-photon detuning $|\Delta/g_a|$. Four colors represent four sizes $L=6$ (black), $7$ (red), $8$ (blue), $9$ (yellow). The corresponding irreducible Hilbert subspaces are $\mathcal{D}_L=399$, $1996$, $9581$, $47692$. Gray dashed lines represent $\langle r\rangle=0.368$ and $0.5295$ in Figs.~\ref{fig2} $(a)-(c)$. The inset in Fig.~\ref{fig2} $(a)$ is $\langle r\rangle$ for HWBC with $L=5$, $6$, $7$. Dashed lines indicate the corresponding GOE value of $\langle \widetilde{D}_{1}\rangle$ and $var(\widetilde{D}_{1})$ in Figs.~\ref{fig2} $(d)-(f)$ and $(g)-(i)$, respectively. The first, second and third column correspond to $|\Delta/g_a|=0$, $5$ and $10$, respectively.}\label{fig2}
\end{figure}

We divide $\epsilon$ into 100 bins of equal width. When the eigenvalues and eigenvectors fall into each bin, mean values ($\langle r\rangle$, $\langle \widetilde{D}_{1}\rangle$) and variances $var(\widetilde{D}_{1})$ as a function of $t/g_a$ can be computed with various atom-photon detuning in Fig.~\ref{fig1}. A quasi-square region of spectral chaos also can be identified with $\Delta/g_a=0$, $t/g_a\simeq0.65$ and $0.1\lesssim\epsilon\lesssim0.9$ in Figs.~\ref{fig1}(a), (d), (g), where $\langle r\rangle \approx 0.5295$, $\langle \widetilde{D}_{1}\rangle$ closes to $1$ and $var(\widetilde{D}_{1})$ is fewer orders of magnitude. On the other hand, other regions of eigenenergy spectrum maybe non-ergodic and cannot be thermalized in the thermodynamic limit. One can find that these three physical quantities exhibit a symmetry about $\epsilon=0.5$. The reason is that, under the chiral operator $\Gamma=\Pi_{j\in even}e^{\pi ia^{\dag}_ja_j}\Pi_{j\in odd}\sigma_j^{z}$, the system has a chiral symmetry in the resonance $\Delta/g_a=0$ for an even lattice site of the PBC\cite{qli1}. From the energy-resolved density plot of $\langle \widetilde{D}_{1}\rangle$, one can find that, the eigenenergy spectra gradually form several energy bands with the increase of $|\Delta/g_a|$ at small scaled tunnelling strength. The region of ergodic spectra fans out as $|\Delta/g_a|$ increases. The eigenenergy spectra no longer have the ergodic region for large $|\Delta/g_a|$. Moreover, The behaviors of $\langle \widetilde{D}_{2}\rangle$ and $\langle \widetilde{D}_{\infty}\rangle$ are similar to that of $\langle \widetilde{D}_{1}\rangle$, thus, we do not discuss these two quantities in this section.

 The eigenenergy average level spacings and eigenstate analyzed are required far from the edges of the spectrum for the applicability of RMT\cite{DAlessio}. Therefore, the low-lying excited states and the states with the highest energies should be excluded and we only chose (middle one-third) a part of the intermediate energy spectrum, which can reflect the ergodic behavior of the considered system. To clarify the influence of the size of Hilbert space on the eigenstates' structure and eigenenergies, we study the mean of $\langle r\rangle$ and $\langle \widetilde{D}_{1}\rangle$ and the $var(\widetilde{D}_{1})$ change with $t/g_a$ and $L$ for the atom-photon detuning $|\Delta/g_a| =0,5,10$ in Fig. \ref{fig2}.

 From Fig.~\ref{fig2} (a), we can see that $\langle r\rangle$ is approximately Wigner-Dyson distribution in the middle coupling strength $t/g_a$ and $|\Delta/g_a| =0$, and this feature is more obvious when the size $L$ is large. Significantly, the average level spacing distributions of odd and even lattice sites are different when the tunneling is large. This is because the even lattice sites have chiral symmetry in the PBC, but odd lattice sites do not. For the case of odd lattice sites, both the labels of the first and the last sites are odd in the tunnelling term, which prevents the Hamiltonian $H$ from satisfying the chiral symmetry under the chiral operator $\Gamma$, and this effect is especially significant for large tunneling strength. It can be seen from the inset of Fig.~\ref{fig2}$(a)$ that under the hard-wall boundary condition (HWBC), both the odd and even lattice sites have chiral symmetry, thus $\langle r\rangle$ is no difference. For $\langle \widetilde{D}_{1}\rangle$ [Fig. \ref{fig2} (d)], when $t/g_a$ is around $1$, it exhibits a wider plateau for increasing $L$ and  shows a extended non-ergodic behavior, but the value of which depends on the size $L$. In addition, the corresponding $var(\widetilde{D}_{1})$ presents minimum value at intermediate range of $t/g_a$ and rises sharply on both sides of the valley [Fig. \ref{fig2} (g). It is noteworthy that the plateau values of $\langle \widetilde{D}_{1}\rangle$ and the valley of $var(\widetilde{D}_{1})$ agree well with those expected value of Eqs. (\ref{H3})-(\ref{H4}) for GOE eigenvectors, shown by dash lines in Figs. \ref{fig2} $(d)$ and $(g)$. Obviously, $\langle \widetilde{D}_{1}\rangle\rightarrow 1$, $var(\widetilde{D}_{1})\rightarrow 0$ and $\langle r\rangle$ approaches $0.5295$, when the size $L\rightarrow\infty$. At the other regions of the parameter $t/g_a$, we are not sure whether $\langle\widetilde{D}_{1}\rangle$ ($\langle r\rangle$) tends to zero (0.386) in the finite-size, but they mismatch the GOE of RMT and these regions are not an ergodic phase, which is similar to Bose Hubbard model\cite{Pausch}.

 Therefore, combining $\langle r\rangle$, $\langle \widetilde{D}_{1}\rangle$ and $var(\widetilde{D}_{1})$, one knows that the intermediate region is a standard ergodic phases and the regions on both sides are non-ergodic phases with $|\Delta/g_a| =0$ in the thermodynamic limit. However, when $|\Delta/g_a|$ is large [The second and third column in Fig. \ref{fig2}], $\langle r\rangle$ of the original ergodic region is close to Poisson distribution with the increase of the size $L$. Meanwhile, the results of $\langle\widetilde{D}_{1}\rangle$ and $var(\widetilde{D}_{1})$ are more and more inconsistent with GOE of RMT and the ergodic region will disappears. We predict that the phase transition occurs from ergodic to non-ergodic cases with increasing atom-photon detuning in the thermodynamic limit.

\begin{figure}[H]
\includegraphics[height=3cm,width=9.3cm]{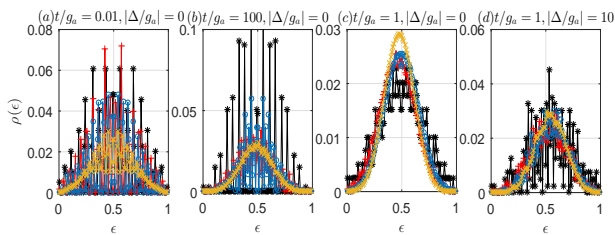}
\caption{(Color online) Density of states coarse-grained over the scaled energy $\epsilon$ for different $t/g_a$ and atom-photon detuning $|\Delta/g_a|$. According to the number of eigenstates, $\epsilon$ is divided into 100 equal bins. The spikes mark the multiplet structure. Four colors and signal represent four sizes $L=6$ (black-star), $7$ (red-plus), $8$ (blue-circle), $9$ (yellow-cross). }\label{fig3}
\end{figure}

 We all know that when $t/g_a$ equals $0$ or $\infty$, the eigenstates are massively degenerate. Thus, the density of states $\rho(\epsilon)=\Sigma_{\alpha}\delta(\epsilon-\epsilon_{\alpha})/\mathcal{D}(L)$ are shown in Fig.~\ref{fig3} to judge whether the average level spacings of Poisson distribution for the large size is caused by energy level quasidegeneracy\cite{Russomanno}. One can see that the density of states shows a series of spikes at $t/g_a=0.01,100$ and $|\Delta/g_a|=0$ for a small system size $L$, but the structure gets diluted into a smooth continuum as increasing the system size. This tendency is also quite obvious at $t/g_a=1$ and $|\Delta/g_a|=10$. Moreover, the density of states is almost a smooth curve at $t/g_a=1$ and $|\Delta/g_a|=0$, even in the case of small system size, which is a Wigner-Dyson distribution and ergodic region. Therefore, the discussions of the ergodicity breaking in a small (large) $t/g_a$ and large $|\Delta/g_a|$ are independent of the spectrum in multiplets.

\begin{figure*}[htb]
\includegraphics[scale=0.90,trim=20 0 0 0]
{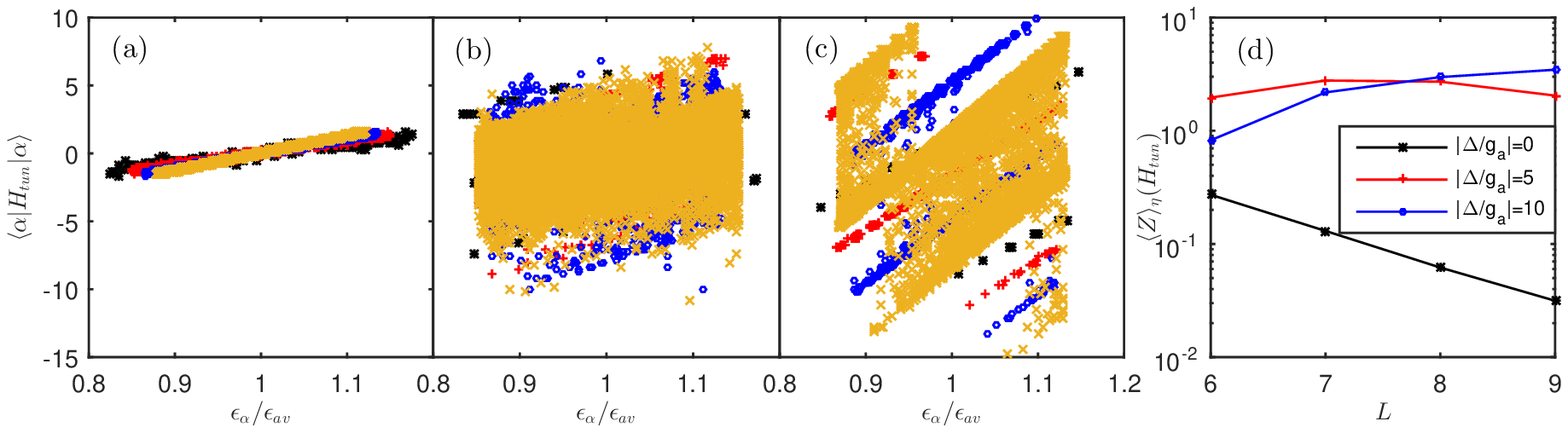}
\includegraphics[scale=0.90,trim=20 0 0 0]
{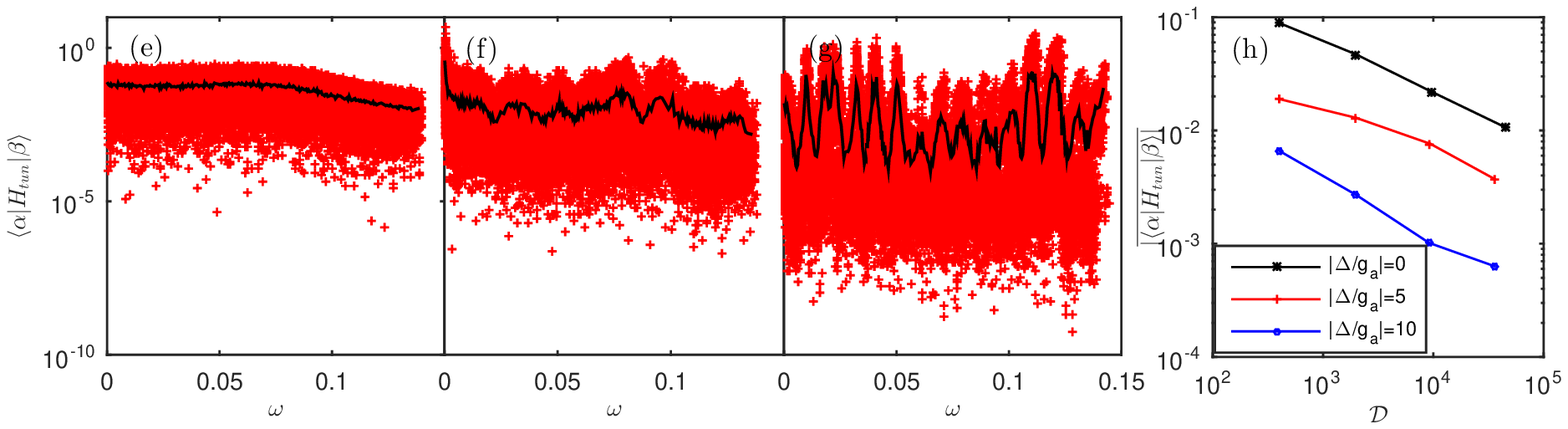}
\caption{(Color online) The diagonal matrix elements of $H_{tun}$  as a function
of the rescaled energy for $|\Delta/g_a|=0$ [(a)], $5$ [(b)], $10$ [(c)] with different sizes $L=6$ (black-star), $7$ (red-plus), $8$ (blue-circle), $9$ (yellow-cross). (d) The mean statistics of eigenstate-to-eigenstate fluctuations in $Z_{\eta}$ vs the system size for the observable $H_{tun}$.  The off-diagonal matrix elements of $H_{tun}$ as a function
of the rescaled energy for $|\Delta/g_a|=0$ [(e)], $5$ [(f)], $10$ [(g)] for 7(red-plus).  The corresponding black line is running averages with a subset length of 100. (d) The average off-diagonal matrix elements $\overline{|\langle\alpha |H_{tun}|\beta\rangle|}$ vs the irreducible Hilbert subspace $\mathcal{D}$. }\label{fig4}
\end{figure*}

In order to understand the thermalization properties of the ergodic and non-ergodic phases, we further study whether the system obeys ETH for different $|\Delta/g_a|$. ETH is often used to describe the mechanism of the quantum thermalization in a generic quantum ergodic system\cite{Srednicki,Srednicki1}. When ETH ansatz is applicable, the diagonal matrix elements' fluctuation and the off-diagonal matrix elements of the observables are exponentially small with an increase of the system size\cite{Deutsch,Srednicki2,Rigol,Linden,Khatami}. In the following, we focus on the middle one-third of eigenstates and verify whether the system meets the validity of the ETH for $t/g_a=1$.  The tunnelling term $H_{tun}$ is chosen, namely the photons kinetic energy, as a selected observable.

First, we study the diagonal part of the observable $\langle\alpha |H_{tun}|\alpha\rangle$. Figs.~\ref{fig4} (a)-(c) show the diagonal matrix elements of $H_{tun}$ in the eigenstates $|\alpha\rangle$ of the irreducible Hilbert subspace as functions of $\epsilon_{\alpha} /\epsilon_{av}$ for different $|\Delta/g_a|$, where $\epsilon_{\alpha}$ is the corresponding  scaled energy eigenvalue of the eigenstate $|\alpha\rangle$ and $\epsilon_{av}=Tr_{\mathcal{D}}\{(H-E_{min})/(E_{max}-E_{min})\}$ is the average energy eigenvalue. The different colored symbols represent different system sizes. We find that the fluctuations of the eigenstate-expectation values decrease with size $L$ in Figs.~\ref{fig4} (a)-(c) and eigenstate-expectation values become smooth functions of the energy density in the thermodynamic limit $L\rightarrow\infty$. However, when $|\Delta/g_a|=10$, the fluctuation of the diagonal matrix element is generally invariant with the size increasing, which is in agreement with the behavior of the integrable system and dissatisfies the diagonal part of the ETH ansatz\cite{Beugeling}. Further, we quantitatively analyse the relation between the eigenstate-to-eigenstate fluctuations and the system's size $L$ in the middle one-third spectrum. A measure of eigenstate-to-eigenstate fluctuations of diagonal expectation values is defined as $z_{\alpha}(H_{tun})=\langle \alpha+1|H_{tun}|\alpha+1\rangle-\langle \alpha|H_{tun}|\alpha\rangle$\cite{Kim}.
In the middle one-third energy window, the mean can be given by
\begin{align}
\langle Z\rangle(H_{tun})=\mathcal{D}^{-1}\sum_{|\alpha\rangle\in\mathcal{D}} |z_{\alpha}(H_{tun})|.\label{H5}
\end{align}
From Fig.~\ref{fig4} (d), one can see that the statistical average of eigenstate-to-eigenstate
fluctuations $\langle Z\rangle_{\alpha}(H_{tun})$ change with size $L$ for different atom-photon detuning $|\Delta/g_a|$. One observe that when $|\Delta/g_a|=0$ the fluctuations exponentially decrease as the system size $L$ increases. As the detuning is relatively large, the fluctuations hardly decrease with size $L$.

 We also briefly discuss the off-diagonal matrix elements $\langle\alpha |H_{tun}|\beta\rangle$ of the tunnelling operator to prove whether the second part of the ETH ansatz is satisfied for different $|\Delta/g_a|$. The eigenstates are limited to a narrow energy window $ (1-\frac{\delta}{2})<\bar{\epsilon}/\epsilon_{av}<(1+\frac{\delta}{2})$, where $\delta=0.01$ is the width of the window and $\bar{\epsilon}=(\epsilon_{\alpha}+\epsilon_{\beta})/2$.
 The running average line becomes rougher with the increase of detuning and the overall fluctuation is larger in the non-ergodic phase than the ergodic one for $\langle\alpha |H_{tun}|\beta\rangle$ versus $\omega=\epsilon_{\alpha}-\epsilon_{\beta}$ in Figs.~\ref{fig4}(e)-(g).
But the average off-diagonal matrix elements of the tunnelling operator $\overline{|\langle\alpha |H_{tun}|\beta\rangle|}$ are exponentially small with different irreducible Hilbert subspaces $\mathcal{D}$ for different $\Delta/g_a$ in Fig.~\ref{fig4}(h). Thus, the difference between ergodic and non-ergodic phases is not obvious for the off-diagonal elements.

 Therefore, the results in Fig.~\ref{fig4} reveal that the observable in the middle one-third spectrum obeys ETH for $|\Delta/g_a|=0$ and violates of ETH ansatz for large $|\Delta/g_a|$.

\section{the ergodicity and phase transition of ground state }
\begin{figure*}[htb]
\includegraphics[scale=0.98,trim=40 0 0 0]
{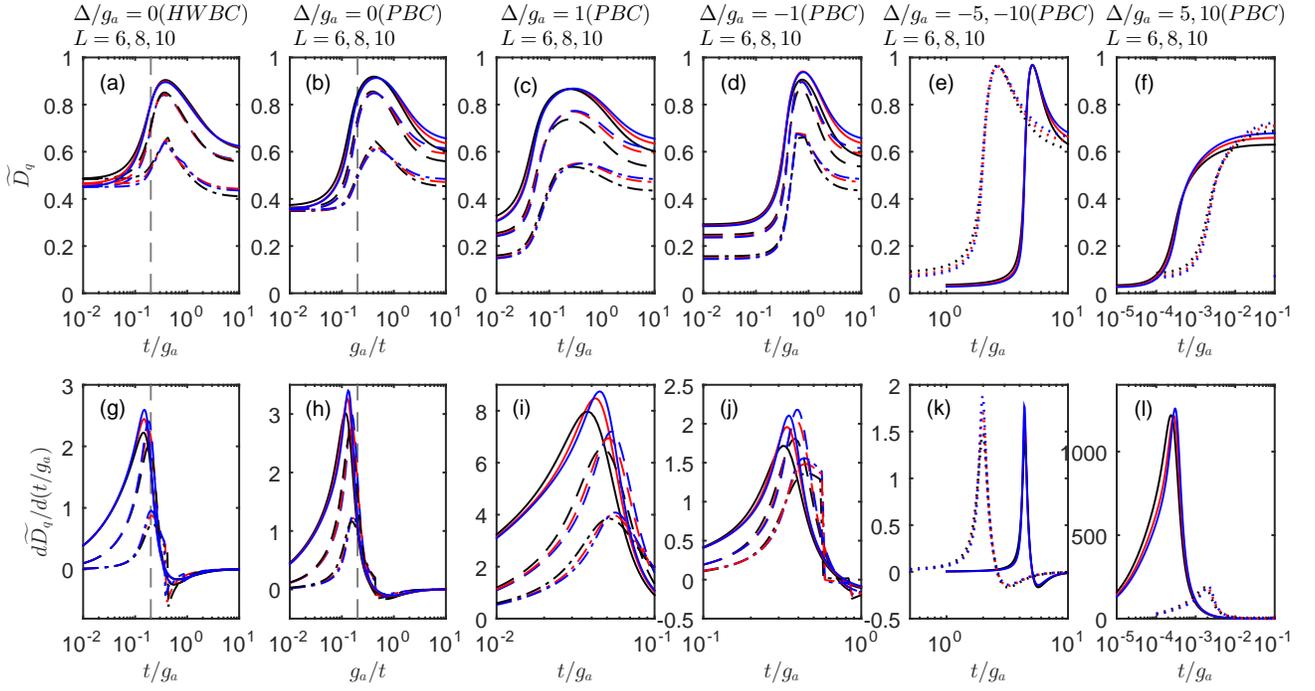}
\caption{(Color online) The finite-size GFDs $\widetilde{D}_{q}$(top) and its derivative $d\widetilde{D}_{q}/d(t/g_a)$ (bottom) as a function of the scaled tunnelling strength $t/g_a$ . From left to right, different subgraphs represent different conditions.  The other columns in the basis of the  product of the each cavity field states and atom states $|\mathbf{n}\rangle$. Black, red and green lines correspond to the size $L$ from small to large. The solid lines are $\widetilde{D}_{1}$, the dashed lines are $\widetilde{D}_{2}$ and the dot-dashed lines are $\widetilde{D}_{\infty}$ in the first four columns. The dotted (solid) line is the $\widetilde{D}_{1}$ for $|\Delta/g_a|=5$ ($10$) in the last two columns. The vertical dotted line marks $t_c/g_a=0.2$.}\label{fig6}
\end{figure*}

In this section, we verify whether GFDs can determine the ground state phase transition and whether the SF and MI phases have a definite many-body multifractality. For integer filling factor $\nu $ and $t/g_a\rightarrow 0$, the ground state is given by $|\Psi(t/g_a=0)\rangle=\prod_i|-,\nu\rangle_i$. Otherwise, when $t/g_a\rightarrow \infty$, we can only consider the tunnelling term of the Hamiltonian $H$ and the ground state is shown by the Fock basis of photons for PBC, $|\Psi(t/g_a=\infty)\rangle=\sum_i^{\mathcal{N}} \sqrt{\frac{N!}{L^N \nu_1!\nu_2!...\nu_L!}}(\prod_{j}|\nu,g\rangle_j)_i$, which is similar to the results of Bose Hubbard model\cite{Lindinger}. One can get GFDs analytically form in the thermodynamic limit for $t/g_a=\infty$ ($D_1=0.941, D_2=0.907, D_{\infty}\approx0.721$, for $\nu=1$)\cite{Lindinger1,Bogomolny}. In other words, the ground state shows multifractality in the Fock basis of photons, which is in an extended non-ergodic phase. Through numerical analysis, one can find that when $t/g_a$ is small, $\widetilde{D}_{q}$ of the ground state is equal to zero, which is a loclalized state. However, when $t/g_a$ is large, the ground state is an extended non-ergodic state with $0<\widetilde{D}_{q}<1$.

In the following, we calculated numerically $\widetilde{D}_1$, $\widetilde{D}_2$ and $\widetilde{D}_{\infty}$ with $t/g_a$ for different conditions. It's well known that the gradient of $\widetilde{D}_{q}$ is the largest value near the phase transition point for the ground state phase transition\cite{Lindinger,Lindinger1}. Therefore, in order to determine the ground state phase transition point, we also plot the derivative of $\widetilde{D}_{q}$ with $t/g_a$.
According to Refs\cite{Lindinger,Lindinger1}, the change of GFDs with $t/g_a$ can characterize the MI-SF phase transition. In Fig.~\ref{fig6}, we draw $\widetilde{D}_{q}$ and $d\widetilde{D}_{q}/d(t/g_a)$ with different $\Delta/g_a$. The vertical dotted line in the subgraphs is the MI-SF phase transition critical point of one-dimensional JCH model, i.e. $(t/g_a)_c\approx0.2$\cite{Rossini,Merin}. The value of $\widetilde{D}_{q}$ gets closer to a small value with $t/g_a\ll1$ as the size $L$ increases, which can be clearly obtained by finite-size analysis. At the large $t/g_a$, the value of $\widetilde{D}_{q}$ shows multifractal behavior and increases with the increasing of size $L$. On the other hand, through the derivative $d\widetilde{D}_{q}/d(t/g_a)$, we can better judge the phase transition critical point. From $d\widetilde{D}_{q}/d(t/g_a)$ versus $t/g_a$ in Figs.~\ref{fig6}(g)-(j), one can find that the absolute value of the maximum derivative of $\widetilde{D}_{q}$ approaches the critical point with the increase of size $L$, which is independent of the boundary conditions. Under different boundary conditions, a slight distinction appears with the increase of size $L$, and $|d\widetilde{D}_{q}/d(t/g_a)|_{max}$ may approach the phase transition critical point in another direction, especially for $\widetilde{D}_{\infty}$. Thus, $d\widetilde{D}_{q}/d(t/g_a)$ is a good quantity to judge the MI-SF phase transition.

 Furthermore, we can determine the range of phase transition critical point $t_c/g_a$ in the case of different filling factor $\nu$ and detuning $\Delta/g_a$ by two directions of the maximum value of $|d\widetilde{D}_{q}/d(t/g_a)|$ approaching the phase transition critical point under PBC and HWBC. For example, we find that for $\Delta/g_a=1$, $t_c/g_a\in(0.06,0.07)$, for $\Delta/g_a=-1$, $t_c/g_a\in(0.39,0.45)$ in Figs. \ref{fig6}(i) and \ref{fig6} (j). Interestingly, the critical point $t_c/g_a$ gradually moves to the right (left) with the increase (decrease) of $\Delta/g_a$ and $\widetilde{D}_{q}$ tends to zero with $t/g_a$ close to zero, as $|\Delta/g_a|$ gets bigger and bigger. In detail, we also found that $\widetilde{D}_{q}$ changes suddenly in a certain value $t/g_a$ especially for $\widetilde{D}_{\infty }$, which makes $\widetilde{D}_{q}$ sometimes unsuitable for describing MI-SF phase transitions. The result of the mutation may be that we consider the subspace of the center-of-mass quasimomentum $Q=0$ in the translational symmetry. However, we find that, when $\Delta/g_a= -1$ or in HWBC, the change of $\widetilde{D}_{\infty }$ is discontinuous even in the whole Hilbert space.  The another result is the inhomogeneous variation of the largest term $|\psi_\alpha|$ in the finite-size. From Fig.~\ref{fig6}, one can draw a conclusion that $\widetilde{D}_{\infty}$, as a function of $t/g_a$, is a smooth curve in the thermodynamic limit. What's particularly interesting is the positive and negative signs of detuning have an opposite effect on the ground state phase transition, which is different from the results of the excited states. Through finite-size analysis, a strict ergodic SF phase maybe not emerged at any atom-photon detuning in the thermodynamic limit for the ground state, which should be further verified by an experiment of the 1D JCH system in the future.

\section{Conclusion and discussion}
In brief, using exact diagonalization, we have provided an integral picture of the ergodic and non-ergodic phases of the one dimensional JCH model for the excited state on the basis of $|\mathbf{n}\rangle$. Through the average level spacings, the GFDs and the GFDs' fluctuations, we can distinguish the ergodic and non-ergodic phases. The ergodic phases of eigenvectors exist in the thermodynamic limit and are well described by RMT. The ergodic phase is transformed into an extended non-ergodic phase with large atom-photon detuning. We also find that the non-ergodic phase closes to Poisson distribution and violates the EHT, especially in the case of large atom-photon detuning. As for the ground state, we describe the characteristics of many-body multifractality for different atom-photon detuning. The rate of GFDs' change can describe the MI-SF phase transition and the extended non-ergodic MI (SF) phase can be achieved. We also find that the phase transition critical point can be identified in a very small range of parameters with large size and different boundary conditions.

\begin{acknowledgments}
This work was supported by National Natural Science Foundation of China (Grants No. 11874190, No. 61835013 and No. 12047501). Support was also provided by Supercomputing Center of Lanzhou University.
\end{acknowledgments}

\end{document}